\documentclass[conference]{IEEEtran}
\usepackage[latin9]{inputenc}
\usepackage{amsmath}
\usepackage{amsthm}
\usepackage{amssymb}
\usepackage{graphicx}

\makeatletter
\theoremstyle{plain}
\newtheorem{thm}{\protect\theoremname}
\theoremstyle{plain}
\newtheorem{prop}[thm]{\protect\propositionname}

\IEEEoverridecommandlockouts
\usepackage{cite}
\usepackage{amsfonts}\usepackage{algorithmic}
\usepackage{textcomp}
\usepackage{xcolor}
\def\BibTeX{{\rm B\kern-.05em{\sc i\kern-.025em b}\kern-.08em
    T\kern-.1667em\lower.7ex\hbox{E}\kern-.125emX}}
\usepackage[acronym]{glossaries}
\newcommand{\newac}{\newacronym}
\newcommand{\ac}{\gls}

\makeglossaries
\newac{isac}{ISAC}{sensing and communication}
\newac{an}{AN}{artificial noise}
\newac{csi}{CSI}{channel state information}
\newac{csit}{CSIT}{channel state information at transmitter}
\newac{bme}{BME}{beampattern matching error}
\newac{ula}{ULA}{uniform linear array}
\newac{cu}{CU}{communication user}
\newac{sinr}{SINR}{signal-to-interference-plus-noise ratio}
\newac{sdr}{SDR}{semi-definite relaxation}
\newac{1d}{1D}{one-dimensional}
\newac{zf}{ZF}{zero-forcing}
\newac{aod}{AoD}{angle of departure}
\newac{los}{LoS}{line-of-sight} 
\newac{nlos}{NLoS}{non-line-of-sight} 
\newac{bs}{BS}{base station}
\newac{bss}{BSs}{base stations}
\newac{cus}{CUs}{communication users}
\newac{snr}{SNR}{signal-to-noise ratio}
\newac{evd}{EVD}{eigenvalue decomposition}
\newac{qsdp}{QSDP}{quadratic semidefinite programing}
\newac{dof}{DoF}{degrees-of-freedom}
\newac{soc}{SOC}{second order cone}
\newac{cscg}{CSCG}{circularly symmetric complex Gaussian}
\newac{crb}{CRB}{Cram{\'e}r-Rao bound}

\providecommand{\propositionname}{Proposition}
\providecommand{\theoremname}{Theorem}

\makeatother

\providecommand{\propositionname}{Proposition}
\providecommand{\theoremname}{Theorem}

\begin{document}
\title{Secure Communications in Near-Filed ISCAP Systems with Extremely Large-Scale
Antenna Arrays{\Large{}\vspace{-0.3cm}} }
\author{\IEEEauthorblockN{{\Large{}(}\textit{\Large{}Invited paper}{\Large{})\vspace{+0.1cm}}}\IEEEauthorblockN{Zixiang Ren$^{1,2}$, Siyao Zhang$^{2}$, Xinmin Li$^{4}$, Ling Qiu$^{1}$,
Jie Xu$^{3,2}$, and {\large{}Derrick Wing Kwan Ng}$^{5}$}\IEEEauthorblockA{$^{1}$Key Laboratory of Wireless-Optical Communications, Chinese
Academy of Sciences,}\IEEEauthorblockA{School of Information Science and Technology, University of Science
and Technology of China}\IEEEauthorblockA{$^{2}$Future Network of Intelligence Institute, The Chinese University
of Hong Kong, Shenzhen}\IEEEauthorblockA{$^{3}$School of Science and Engineering, The Chinese University of
Hong Kong, Shenzhen}\IEEEauthorblockA{$^{4}$College of Computer Science, Chengdu University, Chengdu}
\IEEEauthorblockA{$^{5}$School of Electrical Engineering and Telecommunications, University
of New South Wales} \IEEEauthorblockA{E-mail: rzx66@mail.ustc.edu.cn, zsy@mails.swust.edu.cn, lxm\_edu@126.com, }
\IEEEauthorblockA{lqiu@ustc.edu.cn, xujie@cuhk.edu.cn, w.k.ng@unsw.edu.au{\Large{}\vspace{-0.5cm}}}}
\maketitle
\begin{abstract}
This paper investigates secure communications in a near-field multi-functional
integrated sensing, communication, and powering (ISCAP) system with
an extremely large-scale antenna arrays (ELAA) equipped at the base
station (BS). In this system, the BS sends confidential messages to
a single communication user (CU), and at the same time wirelessly
senses a point target and charges multiple energy receivers (ERs).
It is assumed that the ERs and the sensing target are potential eavesdroppers
that may attempt to intercept the confidential messages intended for
the CU. We consider the joint transmit beamforming design to support
secure communications while ensuring the sensing and powering requirements.
In particular, the BS transmits dedicated sensing/energy beams in
addition to the information beam, which also play the role of artificial
noise (AN) for effectively jamming potential eavesdroppers. Building
upon this, we maximize the secrecy rate at the CU, subject to the
maximum \ac{crb} constraints for target sensing and the minimum
harvested energy constraints for the ERs. Although the formulated
joint beamforming problem is non-convex and challenging to solve,
we acquire the optimal solution via the semi-definite relaxation (SDR)
and fractional programming techniques together with a one-dimensional
(1D) search. Subsequently, we present two alternative designs based
on zero-forcing (ZF) beamforming and maximum ratio transmission (MRT),
respectively. Finally, our numerical results show that our proposed
approaches exploit both the distance-domain resolution of near-field
ELAA and the joint beamforming design for enhancing secure communication
performance while ensuring the sensing and powering requirements in
ISCAP, especially when the CU and the target and ER eavesdroppers
are located at the same angle (but different distances) with respect
to the BS. 
\end{abstract}

\begin{IEEEkeywords}
Integrated sensing, communication, and powering (ISCAP), secure communications,
extremely large-scale antenna array, near-field beamforming, non-convex
optimization.{\Large{}\vspace{-0.4cm}} 
\end{IEEEkeywords}

\section{Introduction\vspace{-0.2cm} }

Integrated sensing and communication (ISAC) and wireless information
and power transfer (WIPT) have emerged as promising technologies for
enabling future sixth-generation (6G) wireless networks, in which
the radio signals conventionally adopted for wireless communications
are reused for the dual roles of environmental sensing and wireless
power transfer (WPT), respectively \cite{liu2022integrated,clerckx2018fundamentals,tong20226g,hua2023optimal}.
With their independent advancements, integrated sensing, communication,
and powering (ISCAP) unifying ISAC and WIPT has recently attracted
growing research interests, which transforms 6G into a new multi-functional
wireless network amalgamating communication, sensing, and WPT functionalities,
thereby achieving synergy and mutual benefits among these essential
functions \cite{chen2024isac}.

Despite the potential benefits, the emergence of ISCAP system introduces
novel data security threats for wireless networks. Due to the involvements
of new sensing and WPT functionalities, the radio signal beams need
to be steered toward sensing targets and energy receivers (ERs). This,
however, may lead to severe information leakage if they are potential
information eavesdroppers. Therefore, it is important but challenging
to provide secure communications while preserving sensing and WPT
requirements. To address this security concern, employing dedicated
sensing/energy signals as artificial noise (AN) is a promising solution.
In this approach, dedicated signal beams can be transmitted jointly
with the information signal beams for offering full degrees of freedom
to enhance sensing and WPT performance, which can also serve as AN
to confuse potential eavesdroppers. While the joint information and
energy/sensing/AN beamforming design has been investigated in ISAC
and WIPT systems independently \cite{liu2014secrecy,su2020secure,ren2023robust},
how to properly design the joint beamforming in ISCAP for efficiently
balancing the performance tradeoff among secure communication, target
sensing, and multiuser WPT has not been well addressed in the literature
yet.

On the other hand, extremely large-scale antenna array (ELAA) is an
evolutionary technology in 6G, which provides significantly enhanced
beamforming gains by increasing the number of antennas at the base
station (BS) an order of magnitude larger than the fifth-generation
(5G) counterpart \cite{lu2021communicating}. In this case, the conventional
designs based on far-field channel properties with planar wavefront
do not hold, and new design approaches based on near-field channels
with spherical wavefront are desirable \cite{cui2022near}. More specifically,
with the spherical wavefront property, the traditional far-field beam
steering evolves into near-field beam focusing \cite{zhang2022beam},
which enables transmitted signal energy to be concentrated on desired
areas in both angular and distance domains concurrently, thereby improving
desired communication signal power and reducing information leakage,
enhancing power transfer efficiency, and achieving accurate target
localization in both angular and distance domains \cite{wang2023near}.
It is thus envisioned that leveraging ELAA in ISCAP systems holds
significant potential to enhance secure communication, target sensing,
and WPT performances simultaneously.

This paper explores secure communications in a near-field multi-functional
ISCAP system with one single CU, one sensing target, and multiple
ERs. The sensing target and ERs act as potential eavesdroppers attempting
to intercept the confidential message intended for the CU. To begin
with, we formulate a joint information and sensing/energy/AN beamforming
problem, with the objective of maximizing the secrecy rate subject
to sensing \ac{crb} constraints for target parameters estimation
and power harvesting constraints for ERs. Despite the non-convex nature
of the formulated joint beamforming problem, we obtain the optimal
solution by exploiting semidefinite relaxation (SDR) and fractional
programming techniques together with a one-dimensional (1D) search.
Furthermore, we present two alternative designs based on zero-forcing
(ZF) beamforming and maximum ratio transmission (MRT), respectively.
Finally, numerical results are provided to demonstrate the effectiveness
of our proposed methods. It is shown that our proposed designs outperform
other schemes by exploiting both the distance-domain resolution of
near-field ELAA and the joint beamforming design for enhancing the
secure communication performance while ensuring the sensing and powering
requirements in ISCAP, especially when the CU and the target and ER
eavesdroppers are located at an identical angle but different distances
with respect to the BS. 

\textit{Notations}: Throughout this paper, vectors and matrices are
denoted by bold lower- and upper-case letters, respectively. $\mathbb{C}^{N\times M}$
denotes the space of $N\times M$ matrices with complex entries. For
a square matrix $\boldsymbol{A}$, $\textrm{Tr}(\boldsymbol{A})$
denotes its trace and $\boldsymbol{A}\succeq\boldsymbol{0}$ means
that $\boldsymbol{A}$ is positive semi-definite. For a complex arbitrary-size
matrix $\boldsymbol{B}$, $\textrm{rank}(\boldsymbol{B})$, $\boldsymbol{B}^{T}$,
$\boldsymbol{B}^{H}$, and denote its rank, transpose, and complex
conjugate, respectively. $\mathbb{E}(\cdot)$ denotes the stochastic
expectation, $\|\cdot\|$ denotes the Euclidean norm of a vector,
and $\mathcal{CN}(\boldsymbol{x},\boldsymbol{Y})$ denotes the \ac{cscg}
random distribution with mean vector $\boldsymbol{x}$ and covariance
matrix $\boldsymbol{Y}$. $(x)^{+}\triangleq\max(x,0)$. $\frac{\partial}{\partial}(\cdot)$
denotes the partial derivative operator. $\mathrm{vec}(\cdot)$ denotes
the vectorization operator.

\section{System Model and Problem Formulation\vspace{-0.1cm} }

\begin{figure}
\vspace{-0.6cm} \includegraphics[scale=0.24]{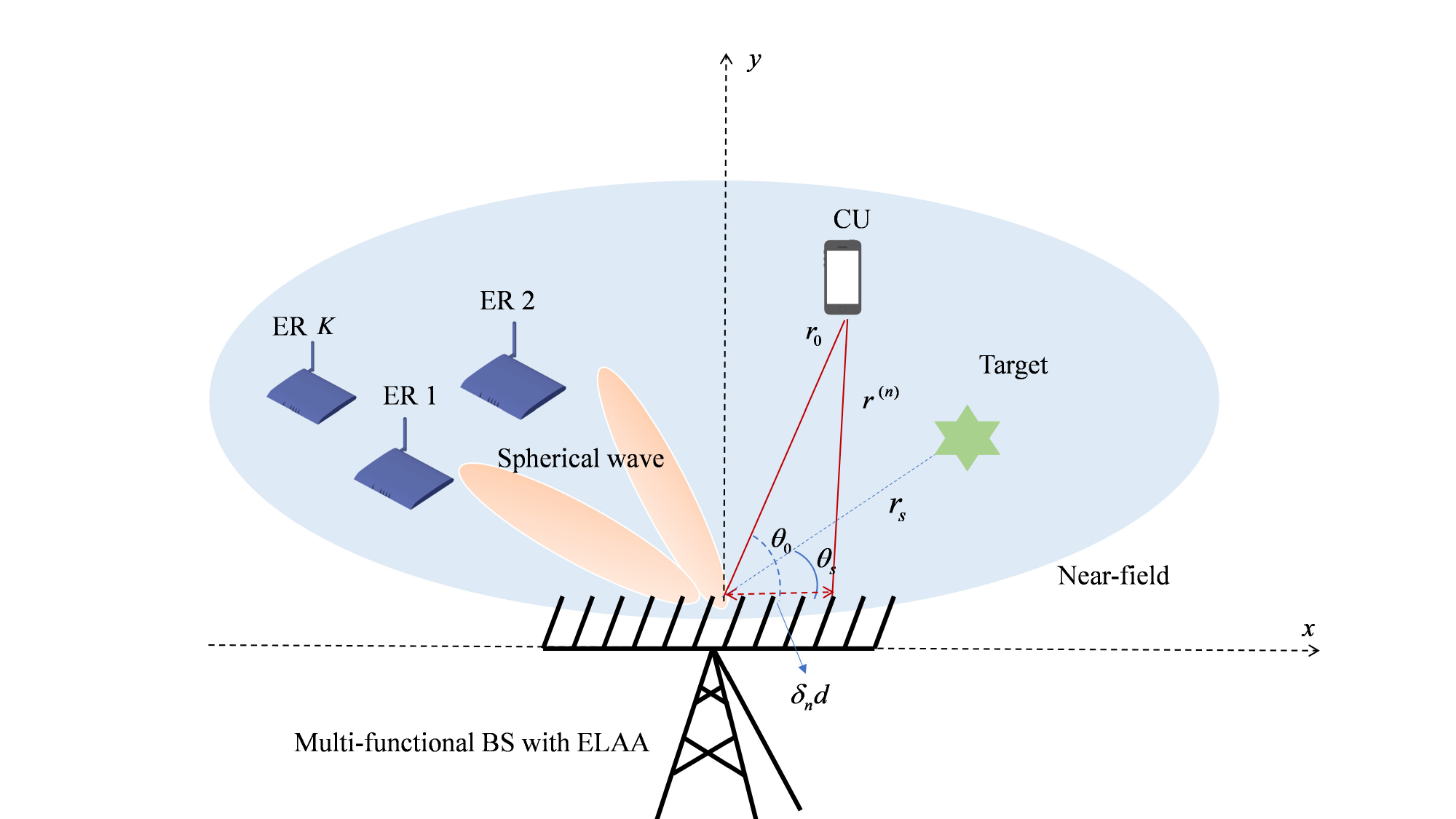}\centering\caption{Illustration of the considered ISCAP system.}
\vspace{-0.4cm} 
\end{figure}

This paper considers a narrowband ISCAP system as shown in Fig. 1,
which compromises a multi-functional BS, one sensing target, $K$
single-antenna ERs, and a single-antenna CU. We assume that the BS
is equipped with an $N$-antenna uniform linear array (ULA) with adjustment
antenna spacing $d$. As a result, the aperture of this antenna array
is $D=(N-1)d$. Let $\lambda$ denote the wavelength of the narrowband
system. We assume that the CU, ERs, and the sensing target are located
in the near-field region of the BS, i.e., their distances from the
BS are less than the Rayleigh distance $\frac{2D^{2}}{\lambda}$ \cite{cui2022near}.
In this scenario, the BS transmits confidential messages to the CU
while simultaneously delivering power to $K$ ERs and conducting target
localization for the sensing target. It is also assumed that the $K$
ERs and the sensing target are potential eavesdroppers that may attempt
to intercept the confidential messages for the CU. Let $\mathcal{K}_{\mathrm{ER}}\overset{\triangle}{=}\{1,2,\dots,K\}$
denote the set of all $K$ ERs and $\mathcal{K}_{\mathrm{EAV}}=\mathcal{K}_{\mathrm{ER}}\cup\{K+1\}$
denote the set of potential eavesdroppers, in which $k=K+1$ represents
the target.

First, we present the joint information and energy/sensing/AN beamforming
design for secure ISCAP. We assume that the BS utilizes transmit beamforming
to transmit the confidential message $s_{0}(t)\in\mathbb{C}$ to the
CU, where $s_{0}(t)$ is a CSCG random variable with zero mean and
unit variance, i.e., $s_{0}(t)\sim\mathcal{CN}(0,1)$, with $t\in\{1,\dots,T\}$
denoting the symbol index. We adopt $\boldsymbol{w}_{0}\in\mathbb{C}^{N\times1}$
to denote the transmit information beamforming vector. In addition
to the information signal $s_{0}(t)$, the BS also transmits dedicated
signals $\boldsymbol{s}_{1}(t)\in\mathbb{C}^{N\times1}$ that play
the triple roles of energy signals, sensing signals, and AN to facilitate
target sensing and energy transmission and to confuse the potential
eavesdroppers. We assume that $\boldsymbol{s}_{1}(t)$ is independent
from $s_{0}(t)$ and is a CSCG random vector with zero mean and covariance
$\boldsymbol{S}=\mathbb{E}(\boldsymbol{s}_{1}(t)\boldsymbol{s}_{1}^{H}(t))\succeq\boldsymbol{0},\textrm{i.e}.,$
$\boldsymbol{s}_{1}(t)\sim\mathcal{CN}(\boldsymbol{0},\boldsymbol{S})$.
We assume that $s_{0}(t)$ and $\boldsymbol{s}_{1}(t)$ are statistically
independent across different symbols, $\forall t\in\{1,\dots,T\}$.
As a result, the transmitted signal by the BS is expressed as\vspace{-0.25cm}
\begin{equation}
\boldsymbol{x}(t)=\boldsymbol{w}_{0}s_{0}(t)+\boldsymbol{s}_{1}(t).
\end{equation}
Consequently, the transmit covariance matrix of $\boldsymbol{x}(t)$
is\vspace{-0.25cm} 
\begin{equation}
\boldsymbol{R}_{x}=\mathbb{E}(\boldsymbol{x}(t)\boldsymbol{x}^{H}(t))=\boldsymbol{W}+\boldsymbol{S},
\end{equation}
where $\boldsymbol{W}=\boldsymbol{w}_{0}\boldsymbol{w}_{0}^{H}$ with
$\boldsymbol{W}\succeq\boldsymbol{0}$ and $\textrm{rank}(\boldsymbol{W})\leq1$.
We consider that the BS is subject to a maximum transmit power budget
$P$. In this case, we have\vspace{-0.25cm} 
\begin{equation}
\mathrm{Tr}(\boldsymbol{W}+\boldsymbol{S})\leq P.\label{eq:power constraint}
\end{equation}

Then, we introduce the near-field channel model. Let $\boldsymbol{g}_{0}\in\mathbb{C}^{N\times1}$
denote the channel vector between the BS and the CU. Let $\boldsymbol{g}_{k}\in\mathbb{C}^{N\times1}$
denote the eavesdropping channel vector between the BS and potential
eavesdropper $k\in\mathcal{K}_{\mathrm{EAV}}$. Without loss of generality,
we suppose that the ULA is oriented along the x-axis, with the origin
being the midpoint. Accordingly, the Cartesian coordinate of its $n$-th
antenna element is $(0,\delta_{n}d)$, where $\delta_{n}=\frac{2n-N+1}{2}$,
$n\in\{0,\dots,N-1\}$. Consider a particular point $(r,\theta)$
in polar coordinates, the distance between the $n$-th element and
the point is given as \cite{cui2022channel}\vspace{-0.25cm} 
\begin{equation}
r^{(n)}=\sqrt{r^{2}+(\delta_{n}d)^{2}-2\delta_{n}dr\cos\theta}.
\end{equation}
As a result, the near-field steering vector is given by\vspace{-0.25cm}
\begin{equation}
\boldsymbol{a}(\theta,r)=\frac{1}{\sqrt{N}}[e^{-j\frac{2\pi}{\lambda}(r^{(0)}-r)},\dots,e^{-j\frac{2\pi}{\lambda}(r^{(N-1)}-r)}]^{T}.
\end{equation}
It is assumed that the near-field channels $\boldsymbol{g}_{k}$'s
each consist of one line-of-sight (LoS) path and $J_{k}\ge0$ scattering
or non-line-of-sight (NLoS) paths, $k\in\{0\}\cup\mathcal{K}_{\mathrm{EAV}}$.
Let $(r_{0},\theta_{0})$ denote the polar coordinate of the CU and
$(r_{k},\theta_{k})$ denote polar coordinate of eavesdropper $k$.
The LoS channel vector for the CU or eavesdropper $k$ in the near-field
region is given as\vspace{-0.25cm} 
\begin{equation}
\boldsymbol{g}_{k}^{\mathrm{LoS}}=\alpha_{k}\boldsymbol{a}(\theta_{k},r_{k}),
\end{equation}
where $|\alpha_{k}|=\frac{c}{4\pi fr_{k}}$ denotes the complex path
gain of the LoS path. Let $\theta_{k}^{j}$ and $r_{k}^{j}$ denote
the angle and distance of the $j$-th path, respectively. Thus, the
NLoS channel component can be modeled as\vspace{-0.25cm} 
\begin{equation}
\boldsymbol{g}_{k}^{\mathrm{NLoS}}=\sum_{j=1}^{J_{0}}\alpha_{k}^{j}\boldsymbol{a}(\theta_{k}^{j},r_{k}^{j}),
\end{equation}
where $\alpha_{k}^{j}\in\mathbb{C}$ represents the complex path gain.
Consequently, the near-field channel between the BS and the CU or
the eavesdropper $k$ is modeled as\vspace{-0.25cm} 
\begin{equation}
\boldsymbol{g}_{k}=\boldsymbol{g}_{k}^{\mathrm{LoS}}+\boldsymbol{g}_{k}^{\mathrm{NLoS}}.
\end{equation}
We assume that $\boldsymbol{g}_{k}$'s are perfectly known at the
BS to facilitate secure ISCAP design \cite{qu2023near,wang2023near}.

Subsequently, we consider the secure communications model. The received
signal at the CU is expressed as\vspace{-0.25cm} 
\begin{equation}
y_{0}(t)=\boldsymbol{g}_{0}^{H}\boldsymbol{w}_{0}s_{0}(t)+\boldsymbol{g}_{0}^{H}\boldsymbol{s}_{1}(t)+z_{0}(t),\label{eq:Received signal at CU}
\end{equation}
where $z_{0}(t)\sim\mathcal{CN}(0,\sigma_{0}^{2})$ denotes the additive
white Gaussian noise (AWGN) at the CU receiver with $\sigma_{0}^{2}$
denoting the noise power. Based on (\ref{eq:Received signal at CU}),
the received \ac{sinr} at the CU is\vspace{-0.25cm} 
\begin{equation}
\gamma_{0}(\boldsymbol{W},\boldsymbol{S},\boldsymbol{h})=\frac{\boldsymbol{g}_{0}^{H}\boldsymbol{W}\boldsymbol{g}_{0}}{\boldsymbol{g}_{0}^{H}\boldsymbol{S}\boldsymbol{g}_{0}+\sigma_{0}^{2}}.\label{eq:SINR at CU}
\end{equation}
Furthermore, the received signal at eavesdropper $k\in\mathcal{K}_{\mathrm{EAV}}$
is denoted as\vspace{-0.25cm} 
\begin{equation}
y_{k}(t)=\boldsymbol{g}_{k}^{H}\boldsymbol{w}_{0}s_{0}(t)+\boldsymbol{g}_{k}^{H}\boldsymbol{s}_{1}(t)+z_{k}(t),\label{eq:Received signal at target}
\end{equation}
where $z_{k}(t)\sim\mathcal{CN}(0,\sigma_{k}^{2})$ denotes the AWGN
at the receiver of eavesdropper $k\in\mathcal{K}_{\mathrm{EAV}}$
with $\sigma_{k}^{2}$ denoting the noise power. Therefore, the SINR
at eavesdropper $k\in\mathcal{K}_{\mathrm{EAV}}$ is\vspace{-0.25cm}
\begin{equation}
\gamma_{k}(\boldsymbol{W},\boldsymbol{S},\boldsymbol{g}_{k})=\frac{\boldsymbol{g}_{k}^{H}\boldsymbol{W}\boldsymbol{g}_{k}}{\boldsymbol{g}_{k}^{H}\boldsymbol{S}\boldsymbol{g}_{k}+\sigma_{k}^{2}}.
\end{equation}
As such, the achievable secrecy rate at the CU under given $\{\boldsymbol{g}_{k}\}$is
given by\vspace{-0.25cm} 
\begin{equation}
\begin{aligned}R(\boldsymbol{W},\boldsymbol{S}) & =\underset{\mathcal{K}_{\mathrm{EAV}}}{\min}\Big(\log_{2}\big(1+\gamma_{0}(\boldsymbol{W},\boldsymbol{S},\boldsymbol{g}_{0})\big)\\
 & -\log_{2}\big(1+\gamma_{k}(\boldsymbol{W},\boldsymbol{S},\boldsymbol{g}_{k})\big)\Big)^{+}.
\end{aligned}
\label{eq:rate}
\end{equation}

Furthermore, we consider energy harvesting at the ERs. Notice that
each ER can harvest wireless energy from both information and dedicated
signals, the received power at ER $k\in\mathcal{K}_{\mathrm{ER}}$
is given as\vspace{-0.25cm} 
\begin{equation}
E_{k}=\zeta\boldsymbol{g}_{k}^{H}(\boldsymbol{W}+\boldsymbol{S})\boldsymbol{g}_{k},
\end{equation}
where $0\leq\zeta\leq1$ denotes the energy harvesting efficiency\footnote{Notice that here we assume linear energy harvesting efficiency. However,
our proposed designs are readily extended to the case with non-linear
energy harvesting efficiency \cite{clerckx2018fundamentals}. }.

Moreover, we consider near-field target sensing. Let $r_{s}$ and
$\theta_{s}$ denote the distance and the angle of the sensing target
to the origin, respectively. Let $\boldsymbol{X}=[\boldsymbol{x}(1),\boldsymbol{x}(2),\dots,\boldsymbol{x}(T)]\in\mathbb{C}^{N\times T}$
and $\boldsymbol{Y}_{s}\in\mathbb{C}^{N\times T}$ denote the accumulated
transmitted signal and received echo signal over the $T$ time slots.
The received echo signal $\boldsymbol{Y}_{s}$ at the BS is denoted
as\vspace{-0.22cm} 
\begin{equation}
\boldsymbol{Y}_{s}=\beta_{s}\boldsymbol{a}(\theta_{s},r_{s})\boldsymbol{a}^{T}(\theta_{s},r_{s})\boldsymbol{X}+\boldsymbol{Z}_{s},
\end{equation}
where $\beta_{s}\in\mathbb{C}$ denotes the complex round-trip channel
coefficient of target depending on the associated path loss and its
radar cross section (RCS), $\boldsymbol{Z}_{s}\in\mathbb{C}^{N\times T}$
denotes the background noise at the BS receiver (including clutters
or interference) with each entry being a zero-mean CSCG random variable
with variance $\sigma_{s}^{2}$. Then, we vectorize matrix $\boldsymbol{Y}_{s}$
as\vspace{-0.22cm} 
\begin{equation}
\boldsymbol{y}_{s}=\hat{\boldsymbol{x}}+\hat{\boldsymbol{z}},
\end{equation}
where $\hat{\boldsymbol{x}}=\mathrm{vec}(\beta_{s}\boldsymbol{a}(\theta_{s},r_{s})\boldsymbol{a}^{T}(\theta_{s},r_{s})\boldsymbol{X})$
and $\hat{\boldsymbol{z}}=\mathrm{vec}(\boldsymbol{Z}_{s})$. In this
scenario, we aim to localize the target via estimating $r_{s}$ and
$\theta_{s}$. We denote $\boldsymbol{\xi}=[\theta_{s},r_{s},\mathrm{Re}(\beta_{s}),\mathrm{Im}(\beta_{s})]$
as unknown parameters to be estimated. The Fisher information matrix
(FIM) $\boldsymbol{J}_{\xi}$ for estimating $\boldsymbol{\xi}$ is
given as \cite{hua2023mimo}\vspace{-0.22cm} 
\begin{equation}
\boldsymbol{J}_{\xi}[i,j]=\frac{1}{\sigma_{s}^{2}}\mathrm{Re}\Big(\frac{\partial\hat{\boldsymbol{x}}^{H}}{\partial\boldsymbol{\xi}[i]}\frac{\partial\hat{\boldsymbol{x}}}{\partial\boldsymbol{\xi}[j]}\Big),i,j=1,\dots,4.
\end{equation}
The CRB matrix is given by the inverse of the FIM, and its diagonal
elements correspond to the CRB of parameters to be estimated. Let
$\boldsymbol{A}=\boldsymbol{a}(\theta_{s},r_{s})\boldsymbol{a}^{T}(\theta_{s},r_{s})$,
$\dot{\boldsymbol{A}}_{\theta}=\frac{\partial\boldsymbol{A}}{\partial\theta_{s}}$,
and $\dot{\boldsymbol{A}}_{r}=\frac{\partial\boldsymbol{A}}{\partial r_{s}}$.
According to \cite{qu2023near}, the CRB for estimating $\theta_{s}$
is given as\vspace{-0.22cm} 
\begin{equation}
\begin{array}{cl}
 & \mathrm{CRB}(\theta_{s},\boldsymbol{W},\boldsymbol{S})\\
 & =\frac{\sigma_{s}^{2}}{2|\beta_{s}|^{2}T}\frac{\mathrm{tr}(\boldsymbol{A}\boldsymbol{R}_{x}\boldsymbol{A}^{H})}{\mathrm{tr}(\dot{\boldsymbol{A}}_{\theta}\boldsymbol{R}_{x}\dot{\boldsymbol{A}}_{\theta}^{H})\mathrm{tr}(\boldsymbol{A}\boldsymbol{R}_{x}\boldsymbol{A}^{H})-|\mathrm{tr}(\boldsymbol{A}\boldsymbol{R}_{x}\dot{\boldsymbol{A}}_{\theta}^{H})|^{2}}.
\end{array}\label{eq:CRB1_theta}
\end{equation}
Similarly, the CRB for estimating $r_{s}$ is given as\vspace{-0.22cm}
\begin{equation}
\begin{array}{cl}
 & \mathrm{CRB}(r_{s},\boldsymbol{W},\boldsymbol{S})\\
 & =\frac{\sigma_{s}^{2}}{2|\beta_{s}|^{2}T}\frac{\mathrm{tr}(\boldsymbol{A}\boldsymbol{R}_{x}\boldsymbol{A}^{H})}{\mathrm{tr}(\dot{\boldsymbol{A}}_{r}\boldsymbol{R}_{x}\dot{\boldsymbol{A}}_{r}^{H})\mathrm{tr}(\boldsymbol{A}\boldsymbol{R}_{x}\boldsymbol{A}^{H})-|\mathrm{tr}(\boldsymbol{A}\boldsymbol{R}_{x}\dot{\boldsymbol{A}}_{r}^{H})|^{2}}.
\end{array}\label{eq:CRB_r}
\end{equation}

Our objective is to maximize the secrecy rate in \eqref{eq:rate},
by jointly optimizing the transmit information covariance matrix $\boldsymbol{W}$
and the sensing/energy/AN covariance matrix $\boldsymbol{S}$, subject
to the requirements on target sensing and WPT. The secrecy rate maximization
problem is formulated as\vspace{-0.22cm} 
\begin{subequations}
\begin{eqnarray}
\textrm{(P1):} & \underset{\boldsymbol{W},\boldsymbol{S}}{\max} & R(\boldsymbol{W},\boldsymbol{S})\nonumber \\
 & \textrm{s.t.} & \mathrm{CRB}(\theta_{s},\boldsymbol{W},\boldsymbol{S})\leq\Gamma_{\theta},\nonumber \\
 &  & \mathrm{CRB}(r_{s},\boldsymbol{W},\boldsymbol{S})\leq\Gamma_{r},\label{eq:CRB constraint}\\
 &  & \zeta\boldsymbol{g}_{k}^{H}(\boldsymbol{W}+\boldsymbol{S})\boldsymbol{g}_{k}\geq Q,\forall k\in\mathcal{K}_{\mathrm{ER}},\label{eq:powering constraint}\\
 &  & \mathrm{Tr}(\boldsymbol{W}+\boldsymbol{S})\leq P,\label{eq:transmit power}\\
 &  & \boldsymbol{W}\succeq\boldsymbol{0},\boldsymbol{S}\succeq\boldsymbol{0},\label{eq:semidefinite constraint}\\
 &  & \textrm{rank}(\boldsymbol{W})\le1,\label{eq:rank constraint}
\end{eqnarray}
\end{subequations}
where $\Gamma_{\theta}$, $\Gamma_{r}$, and $Q$ denote the given
thresholds for angle estimation, range estimation, and energy harvesting,
respectively. Solving problem (P1) is generally challenging as the
objective function and constraints \eqref{eq:CRB constraint} and
\eqref{eq:rank constraint} are non-convex.

\section{Optimal Solution to Problem (P1)}

This section presents the optimal solution to problem (P1). To reduce
the solution complexity caused by the large dimension of ELAA, we
first restrict the optimization of $\boldsymbol{W}$ and $\boldsymbol{S}$
in the subspace spanned by the sensing, communication, and powering
channels. Then, we propose the optimal solution to the reformulated
problem with reduced dimension.

\subsection{Dimension Reduction}

It is observed that only the signal components lying in the subspaces
spanned by $\boldsymbol{H}=\left[\boldsymbol{g}_{0},\dots,\boldsymbol{g}_{K+1},\boldsymbol{a},\frac{\partial\boldsymbol{a}}{\partial\theta_{s}},\frac{\partial\boldsymbol{a}}{\partial r_{s}}\right]\in\mathbb{C}^{N\times(K+5)}$
contribute to problem (P1). In this case, suppose that the rank of
the accumulated matrix $\boldsymbol{H}$ is $L$, i.e., $\mathrm{rank}\big(\boldsymbol{H}\big)=L\le N$,
and its truncated singular value decomposition (SVD) is\vspace{-0.22cm}
\begin{equation}
\boldsymbol{H}=\boldsymbol{U}\boldsymbol{\Lambda}\boldsymbol{V}^{H},
\end{equation}
where $\boldsymbol{U}\in\mathbb{C}^{N\times J}$ and $\boldsymbol{V}\in\mathbb{C}^{(K+5)\times J}$
collect the left and right singular vectors corresponding to the non-zero
singular values, respectively. In this case, we express the transmit
covariance matrix $\boldsymbol{S}$ and $\boldsymbol{W}$ as $\boldsymbol{S}=\boldsymbol{U}\bar{\boldsymbol{S}_{x}}\boldsymbol{U}^{H}$
and $\boldsymbol{W}=\boldsymbol{U}\bar{\boldsymbol{W}}\boldsymbol{U}^{H}$,
respectively, where $\bar{\boldsymbol{S}}\in\mathbb{C}^{L\times L}$
and $\bar{\boldsymbol{W}}\in\mathbb{C}^{L\times L}$correspond to
the equivalent transmit covariance matrix to be optimized. Let $\bar{\boldsymbol{g}}_{k}=\boldsymbol{U}^{H}\boldsymbol{g}_{k},\forall k\in\{0\}\cup\mathcal{K}_{\mathrm{EAV}}$
denote the projected channels in the subspace. In this case, we reduce
the dimension of optimization variable from $N$ (for $\boldsymbol{S}$)
to $L$ (for $\bar{\boldsymbol{S}}$). Accordingly, we equivalently
reformulate problem (P1) as \vspace{-0.2cm} 
\begin{subequations}
\begin{eqnarray}
\textrm{(P2):} & \underset{\bar{\boldsymbol{W}},\bar{\boldsymbol{S}}}{\max} & R(\bar{\boldsymbol{W}},\bar{\boldsymbol{S}})\nonumber \\
 & \textrm{s.t.} & \mathrm{CRB}(\theta_{s},\bar{\boldsymbol{W}},\bar{\boldsymbol{S}})\leq\Gamma_{\theta},\nonumber \\
 &  & \mathrm{CRB}(r_{s},\bar{\boldsymbol{W}},\bar{\boldsymbol{S}})\leq\Gamma_{r},\label{eq:CRB constraint-1}\\
 &  & \zeta\bar{\boldsymbol{g}}_{k}^{H}(\bar{\boldsymbol{W}}+\bar{\boldsymbol{S}})\bar{\boldsymbol{g}}_{k}\geq Q,\forall k\in\mathcal{K}_{\mathrm{ER}},\label{eq:powering constraint-1}\\
 &  & \mathrm{Tr}(\bar{\boldsymbol{W}}+\bar{\boldsymbol{S}})\leq P,\label{eq:transmit power-1}\\
 &  & \bar{\boldsymbol{W}}\succeq\boldsymbol{0},\bar{\boldsymbol{S}}\succeq\boldsymbol{0},\label{eq:semidefinite constraint-1}\\
 &  & \textrm{rank}(\bar{\boldsymbol{W}})\le1.\label{eq:rank constraint-1}
\end{eqnarray}
\end{subequations}
Notice that $R(\bar{\boldsymbol{W}},\bar{\boldsymbol{S}})$, $\mathrm{CRB}(\theta_{s},\bar{\boldsymbol{W}},\bar{\boldsymbol{S}})$,
and $\mathrm{CRB}(r_{s},\bar{\boldsymbol{W}},\bar{\boldsymbol{S}})$
can be obtained via replacing $\boldsymbol{W}$ and $\boldsymbol{S}$
in \eqref{eq:rate}, \eqref{eq:CRB1_theta}, and \eqref{eq:CRB_r}
with $\boldsymbol{S}=\boldsymbol{U}\bar{\boldsymbol{S}_{x}}\boldsymbol{U}^{H}$,
$\boldsymbol{W}={\boldsymbol{U}}\bar{\boldsymbol{W}}{\boldsymbol{U}}^{H}$.
However, problem (P2) is still difficult to solve due to the non-convexity
of the objective function and the constraints in \eqref{eq:CRB constraint-1}
and \eqref{eq:rank constraint-1}.

\subsection{Optimal Solution to Problem (P2)}

To solve (P2), we first drop the rank constraint in \eqref{eq:rank constraint-1}
to obtain the SDR version of problem (P2) as\vspace{-0.2cm} 
\begin{eqnarray*}
\textrm{(SDR2):} & \underset{\bar{\boldsymbol{W}},\bar{\boldsymbol{S}}}{\max} & R(\bar{\boldsymbol{W}},\bar{\boldsymbol{S}})\\
 & \textrm{s.t.} & \textrm{\eqref{eq:CRB constraint-1}, \eqref{eq:powering constraint-1}, \eqref{eq:transmit power-1}, and \eqref{eq:semidefinite constraint-1}}.
\end{eqnarray*}
We further adopt the Schur component to reformulate the CRB constraint
$\mathrm{CRB}(\theta_{s},\bar{\boldsymbol{W}},\bar{\boldsymbol{S}})\leq\Gamma_{\theta}$
as \cite{liu2021cramer}\vspace{-0.2cm} 

\begin{equation}
\left[\begin{array}{cc}
\big(\mathrm{tr}(\dot{\boldsymbol{A}}_{\theta}\boldsymbol{R}_{x}\dot{\boldsymbol{A}}_{\theta}^{H})-\frac{\sigma_{s}^{2}}{2|\beta_{s}|^{2}T\Gamma_{\theta}}\big) & \mathrm{tr}(\dot{\boldsymbol{A}_{\theta}}\boldsymbol{R}_{x}\boldsymbol{A}^{H})\\
\mathrm{tr}(\boldsymbol{A}\boldsymbol{R}_{x}\dot{\boldsymbol{A}}_{\theta}^{H}) & \mathrm{tr}(\boldsymbol{A}\boldsymbol{R}_{x}\boldsymbol{A}^{H})
\end{array}\right]\succeq\boldsymbol{0},\label{eq:CRB1}
\end{equation}
where $\boldsymbol{R}_{x}=\boldsymbol{U}(\bar{\boldsymbol{W}}+\bar{\boldsymbol{S}})\boldsymbol{U}^{H}$.
Similarly, the CRB constraint $\mathrm{CRB}(r_{s},\bar{\boldsymbol{W}},\bar{\boldsymbol{S}})\leq\Gamma_{r}$
is reformulated as\vspace{-0.2cm} 
\begin{equation}
\left[\begin{array}{cc}
\big(\mathrm{tr}(\dot{\boldsymbol{A}}_{r}\boldsymbol{R}_{x}\dot{\boldsymbol{A}}_{r}^{H})-\frac{\sigma_{s}^{2}}{2|\beta_{s}|^{2}T\Gamma_{r}}\big) & \mathrm{tr}(\dot{\boldsymbol{A}_{r}}\boldsymbol{R}_{x}\boldsymbol{A}^{H})\\
\mathrm{tr}(\boldsymbol{A}\boldsymbol{R}_{x}\dot{\boldsymbol{A}}_{r}^{H}) & \mathrm{tr}(\boldsymbol{A}\boldsymbol{R}_{x}\boldsymbol{A}^{H})
\end{array}\right]\succeq\boldsymbol{0}.\label{eq:CRB2}
\end{equation}

Then, we handle the non-convex objective function. First, we introduce
an auxiliary variable $\gamma_{R}$ as an eavesdropping SINR threshold,
which is a variable to be optimized. As such, we and equivalently
reformulate problem (SDR2) as\vspace{-0.2cm} 
\begin{eqnarray*}
\textrm{(SDR2.1):} & \hspace{-0.2cm}\underset{\bar{\boldsymbol{W}},\bar{\boldsymbol{S}},\gamma_{R}}{\max} & \hspace{-0.2cm}\frac{\bar{\boldsymbol{g}}_{0}^{H}\bar{\boldsymbol{W}}\bar{\boldsymbol{g}}_{0}}{\bar{\boldsymbol{g}}_{0}^{H}\bar{\boldsymbol{S}}\bar{\boldsymbol{g}}_{0}+\sigma_{0}^{2}}\\
 & \hspace{-0.2cm}\textrm{s.t.} & \hspace{-0.5cm}\bar{\boldsymbol{g}}_{k}^{H}\bar{\boldsymbol{W}}\bar{\boldsymbol{g}}_{k}\leq\gamma_{R}(\bar{\boldsymbol{g}}_{k}^{H}\bar{\boldsymbol{S}}\bar{\boldsymbol{g}}_{k}+\sigma_{k}^{2}),\forall k\in\mathcal{K}_{\mathrm{EAV}},\\
 &  & \hspace{-0.2cm}\textrm{\eqref{eq:CRB1}, \eqref{eq:CRB2}, \eqref{eq:powering constraint-1}, \eqref{eq:transmit power-1}, and \eqref{eq:semidefinite constraint-1}}.
\end{eqnarray*}
It is worth noting that the objective function is still non-convex.
We introduce a variable $\xi>0$ and adopt the Charnes-Cooper transformation
\cite{shen2018fractional} by defining $\hat{\boldsymbol{W}}=\xi\bar{\boldsymbol{W}}$
and $\hat{\boldsymbol{S}}=\xi\bar{\boldsymbol{S}}$. The CRB constraints
in $\eqref{eq:CRB1},\eqref{eq:CRB2}$ are equivalently reformulated
as\vspace{-0.2cm} 
\begin{equation}
\left[\begin{array}{cc}
\big(\mathrm{tr}(\dot{\boldsymbol{A}}_{\theta}\hat{\boldsymbol{R}_{x}}\dot{\boldsymbol{A}}_{\theta}^{H})-\frac{\xi\sigma_{s}^{2}}{2|\beta_{s}|^{2}T\Gamma_{\theta}}\big) & \mathrm{tr}(\dot{\boldsymbol{A}_{\theta}}\hat{\boldsymbol{R}_{x}}\boldsymbol{A}^{H})\\
\mathrm{tr}(\boldsymbol{A}\hat{\boldsymbol{R}_{x}}\dot{\boldsymbol{A}}_{\theta}^{H}) & \mathrm{tr}(\boldsymbol{A}\hat{\boldsymbol{R}_{x}}\boldsymbol{A}^{H})
\end{array}\right]\succeq\boldsymbol{0},\label{eq:CRB1-1}
\end{equation}
\begin{equation}
\left[\begin{array}{cc}
\big(\mathrm{tr}(\dot{\boldsymbol{A}}_{r}\hat{\boldsymbol{R}_{x}}\dot{\boldsymbol{A}}_{r}^{H})-\frac{\xi\sigma_{s}^{2}}{2|\beta_{s}|^{2}T\Gamma_{r}}\big) & \mathrm{tr}(\dot{\boldsymbol{A}_{r}}\hat{\boldsymbol{R}_{x}}\boldsymbol{A}^{H})\\
\mathrm{tr}(\boldsymbol{A}\hat{\boldsymbol{R}_{x}}\dot{\boldsymbol{A}}_{r}^{H}) & \mathrm{tr}(\boldsymbol{A}\hat{\boldsymbol{R}_{x}}\boldsymbol{A}^{H})
\end{array}\right]\succeq\boldsymbol{0},\label{eq:CRB2-1}
\end{equation}
where $\hat{\boldsymbol{R}_{x}}=\boldsymbol{U}(\hat{\boldsymbol{W}}+\hat{\boldsymbol{S}})\boldsymbol{U}^{H}$.
Problem (SDR2.1) is equivalently reformulated as\vspace{-0.2cm} 
\begin{subequations}
\begin{eqnarray}
\hspace{-0.5cm}\textrm{(SDR2.2):} & \underset{\hat{\boldsymbol{W}},\hat{\boldsymbol{S}},\gamma_{R},\xi>0}{\max} & \bar{\boldsymbol{g}}_{0}^{H}\hat{\boldsymbol{W}}\bar{\boldsymbol{g}}_{0}\nonumber \\
 & \textrm{s.t.} & \bar{\boldsymbol{g}}_{k}^{H}\hat{\boldsymbol{W}}\bar{\boldsymbol{g}}_{k}\leq\gamma_{R}(\bar{\boldsymbol{g}}_{k}^{H}\hat{\boldsymbol{S}}\bar{\boldsymbol{g}}_{k}+\xi\sigma_{k}^{2}),\nonumber \\
 &  & \forall k\in\mathcal{K}_{\mathrm{EAV}},\label{eq:C1}\\
 &  & \bar{\boldsymbol{h}}^{H}\hat{\boldsymbol{S}}\bar{\boldsymbol{h}}+\xi\sigma_{0}^{2}=1,\label{eq:C2}\\
 &  & \hspace{-0.8cm}\zeta\bar{\boldsymbol{g}}_{k}^{H}(\hat{\boldsymbol{W}}+\hat{\boldsymbol{S}})\bar{\boldsymbol{g}}_{k}\geq\xi Q,\forall k\in\mathcal{K}_{\mathrm{ER}},\label{eq:C3}\\
 &  & \mathrm{Tr}(\hat{\boldsymbol{W}}+\hat{\boldsymbol{S}})\leq\xi P,\label{eq:C4}\\
 &  & \hat{\boldsymbol{W}}\succeq\boldsymbol{0},\hat{\boldsymbol{S}}\succeq\boldsymbol{0},\label{eq:C5}\\
 &  & \textrm{\eqref{eq:CRB1-1} and \eqref{eq:CRB2-1}}.\nonumber 
\end{eqnarray}
\end{subequations}
 Notice that for a given threshold $\gamma_{R}$, problem (SDR2.2)
is reduced to the following semi-definite programming (SDP) problem
(SDR2.3) that is solvable via off-the-shelf tools such as CVX \cite{cvx}.
\vspace{-0.2cm} 
\begin{eqnarray*}
\textrm{(SDR2.3):} & \underset{\hat{\boldsymbol{W}},\hat{\boldsymbol{S}},\xi>0}{\max} & \bar{\boldsymbol{g}}_{0}^{H}\hat{\boldsymbol{W}}\bar{\boldsymbol{g}}_{0}\\
 & \textrm{s.t.} & \textrm{\eqref{eq:C1}-\eqref{eq:C5}, \textrm{\eqref{eq:CRB1-1}, and \eqref{eq:CRB2-1}} }
\end{eqnarray*}
 As a result, we optimally solve problem (SDR2.2) via solving (SDR2.3)
optimally together with a 1D search over $\gamma_{R}$. Therefore,
problem (SDR2) is optimally solved. 
\begin{prop}
\textup{Let }$\bar{\boldsymbol{W}}^{\star}$\textup{ and $\bar{\boldsymbol{S}}^{\star}$
denote the obtained optimal solution to problem (SDR2). We can always
construct an equivalent solution }$\bar{\boldsymbol{W}}^{\mathrm{opt}}$\textup{
and $\bar{\boldsymbol{S}}^{\mathrm{opt}}$ in the following, such
that the same objective value in (P2) is achieved with rank(}$\bar{\boldsymbol{W}}^{\mathrm{opt}}$\textup{)
= 1.}\vspace{-0.3cm} \textup{
\begin{equation}
\begin{aligned} & \bar{\boldsymbol{W}}^{\mathrm{opt}}=\frac{\bar{\boldsymbol{W}}^{\star}\bar{\boldsymbol{g}}_{0}\bar{\boldsymbol{g}}_{0}^{H}\bar{\boldsymbol{W}}^{\star}}{\bar{\boldsymbol{g}}_{0}^{H}\bar{\boldsymbol{W}}^{\star}\bar{\boldsymbol{g}}_{0}},\bar{\boldsymbol{S}}^{\mathrm{opt}}=\bar{\boldsymbol{W}}^{\star}+\bar{\boldsymbol{S}}^{\star}-\bar{\boldsymbol{W}}^{\mathrm{opt}}.\end{aligned}
\end{equation}
As a result, the constructed solution of }$\bar{\boldsymbol{W}}^{\mathrm{opt}}$\textup{
and $\bar{\boldsymbol{S}}^{\mathrm{opt}}$ is optimal to problem (P2).}\vspace{-0.15cm} 
\end{prop}
\begin{IEEEproof}
The proof is motivated by the proof technique in \cite{ren2023robust}.
The details are omitted due to page limitation.\vspace{-0.2cm} 
\end{IEEEproof}

\section{Alternative Solutions based on ZF and MRT}

In this section, we propose two alternative designs based on ZF and
MRT principles, respectively.

\subsection{ZF-based Beamforming}

In the ZF-based beamforming design, the information beamforming vector
$\boldsymbol{w}_{0}$ is enforced as $\boldsymbol{g}_{k}^{H}\boldsymbol{w}_{0}=0,\forall k\in\mathcal{K}_{\mathrm{EAV}}$.
Moreover, we restrict the transmit sensing/power/AN covariance $\boldsymbol{S}$
in the null space of communication channel to avoid harmful interference.

Let $\boldsymbol{G}=[\boldsymbol{g}_{1},\boldsymbol{g}_{2},\ldots,\boldsymbol{g}_{K+1}]^{H}$
denote the channel matrix from the BS to all the eavesdroppers, of
which the singular value decomposition (SVD) is\vspace{-0.25cm} 
\begin{equation}
\boldsymbol{G}=\boldsymbol{\boldsymbol{\bar{U}}\bar{\Lambda}}\bar{\boldsymbol{V}}^{H}=\boldsymbol{\boldsymbol{\bar{U}}\Lambda}[\boldsymbol{V}_{1}\boldsymbol{V}_{2}]^{H},
\end{equation}
where $\boldsymbol{\bar{U}}\in\mathbb{C}^{(K+1)\times(K+1)}$ and
$\boldsymbol{V}\in\mathbb{C}^{N\times N}$ are both unitary matrices,
and $\boldsymbol{V}_{1}\in\mathbb{C}^{N\times(K+1)}$ and $\boldsymbol{V}_{2}\in\mathbb{C}^{N\times(N-(K+1))}$
consist of the first $(K+1)$ and and the last $N-(K+1)$ right singular
vectors of $\boldsymbol{G}$, respectively. In order to ensure $\boldsymbol{g}_{k}^{H}\boldsymbol{w}_{0}=0,\forall k\in\mathcal{K}_{\mathrm{EAV}}$,
we set\vspace{-0.24cm} 
\begin{equation}
\boldsymbol{w}_{0}=\boldsymbol{V}_{2}\bar{\boldsymbol{w}}_{0},
\end{equation}
where $\bar{\boldsymbol{w}}_{0}\in\mathbb{C}^{(N-(K+1))\times1}$
denotes the ZF beamforming vector. Here, we set the ZF beamforming
vector along the communication channel $\boldsymbol{V}_{2}^{H}\boldsymbol{g}_{0}$,
i.e.,\vspace{-0.24cm} 
\begin{equation}
\bar{\boldsymbol{w}}_{0}=\sqrt{\bar{p_{0}}}\frac{\boldsymbol{V}_{2}^{H}\boldsymbol{g}_{0}}{\|\boldsymbol{V}_{2}^{H}\boldsymbol{g}_{0}\|},
\end{equation}
where $\bar{p_{0}}$ is the allocated communication transmit power
to be optimized. Let $\boldsymbol{V}_{s}=\boldsymbol{I}-\boldsymbol{g}_{0}\boldsymbol{g}_{0}^{H}/\|\boldsymbol{g}_{0}\|^{2}$.
We set the transmit covariance $\boldsymbol{S}$ in the null space
of communication channel to avoid interference, i.e.,\vspace{-0.23cm}
\begin{equation}
\boldsymbol{S}=\boldsymbol{V}_{s}\boldsymbol{S}_{2}\boldsymbol{V}_{s}^{H},
\end{equation}
where $\boldsymbol{S}_{2}\in\mathbb{C}^{N\times N}$ is the transmit
covariance to be optimized. In this case, the secrecy rate becomes\vspace{-0.24cm}
\begin{equation}
R(\bar{\boldsymbol{w}}_{0})=\log_{2}\big(1+\frac{\bar{p_{0}}}{\|\boldsymbol{V}_{2}^{H}\boldsymbol{g}_{0}\|^{2}\sigma_{0}^{2}}\big).
\end{equation}
As a result, we can maximize the transmit power $\bar{p}_{0}$ to
equivalently maximize the secrecy rate, for which the optimization
problem is formulated as\vspace{-0.25cm} 
\[
\begin{array}[b]{ccl}
\textrm{(P3)}: & \underset{\bar{p_{0}},\boldsymbol{S}_{2}}{\max} & \bar{p_{0}}\\
 & \textrm{s.t.} & \overline{\mathrm{CRB}}(\theta_{s},\bar{p_{0}},\boldsymbol{S}_{2})\leq\Gamma_{\theta},\\
 &  & \overline{\mathrm{CRB}}(r_{s},\bar{p_{0}},\boldsymbol{S}_{2})\leq\Gamma_{r},\\
 &  & \zeta\boldsymbol{g}_{k}^{H}\boldsymbol{V}_{s}\boldsymbol{S}_{2}\boldsymbol{V}_{s}^{H}\boldsymbol{g}_{k}\geq Q,\forall k\in\mathcal{K}_{\mathrm{ER}},\\
 &  & \textrm{Tr}(\boldsymbol{V}_{s}\boldsymbol{S}_{2}\boldsymbol{V}_{s}^{H})+\bar{p_{0}}\leq P,\\
 &  & \boldsymbol{S}_{2}\succeq\boldsymbol{0},p_{0}\geq0,
\end{array}
\]
where $\overline{\mathrm{CRB}}(\theta_{s},\bar{p_{0}},\boldsymbol{S}_{2})$
and $\overline{\mathrm{CRB}}(r_{s},\bar{p_{0}},\boldsymbol{S}_{2})$
are the corresponding CRB expression after variable transformation.
Problem (P3) is a typical SDP that can be easily solved via CVX \cite{cvx}.

\subsection{MRT-based Beamforming}

In the MRT-based beamforming, the BS transmits the information beam
for CU, in addition to one sensing beam for target sensing, and $K$
energy beams each for one ER, in which the beamforming is designed
based on the MRT principle. Let $p_{0}$, $p_{s}$, and $\{p_{k}\}$
denote the allocated power dedicated for CU, target, and ERs, respectively.
As such, we set the transmit covariance $\boldsymbol{W}$ and $\boldsymbol{S}$
as\vspace{-0.25cm} 
\begin{equation}
\begin{alignedat}{1} & \boldsymbol{w}_{0}=\sqrt{p_{0}}\boldsymbol{g}_{0}/\|\boldsymbol{g}_{0}\|,\boldsymbol{W}=p_{0}\boldsymbol{g}_{0}\boldsymbol{g}_{0}^{H}/\|\boldsymbol{g}_{0}\|^{2},\\
 & \boldsymbol{S}=\sum_{k=1}^{K}p_{k}\boldsymbol{g}_{k}\boldsymbol{g}_{k}^{H}/\|\boldsymbol{g}_{k}\|^{2}+p_{s}\boldsymbol{a}(\theta_{s},r_{s})\boldsymbol{a}^{H}(\theta_{s},r_{s}).
\end{alignedat}
\label{eq:eqMRT}
\end{equation}
As a result, the optimization of $\boldsymbol{W}$ and $\boldsymbol{S}$
is reduced to the power allocation optimization of $p_{0}$, $p_{s}$,
and $\{p_{k}\}$. In this case, the secrecy rate maximization problem
with MRT-based beamforming is\vspace{-0.25cm} 
\begin{eqnarray*}
\textrm{(P4):} & \hspace{-0.2cm}\underset{p_{0},p_{s},\{p_{k}\}}{\max} & R(p_{0},p_{s},\{p_{k}\})\\
 & \hspace{-0.2cm}\textrm{s.t.} & \widehat{\mathrm{CRB}}(\theta_{s},p_{0},p_{s},\{p_{k}\})\leq\Gamma_{\theta},\\
 &  & \widehat{\mathrm{CRB}}(r_{s},p_{0},p_{s},\{p_{k}\})\leq\Gamma_{r},\\
 &  & \hspace{-0.4cm}\zeta\boldsymbol{g}_{k}^{H}(p_{0}\boldsymbol{h}\boldsymbol{h}^{H}/\|\boldsymbol{h}\|^{H}+\boldsymbol{S})\boldsymbol{g}_{k}\geq Q,\forall k\in\mathcal{K}_{\mathrm{ER}},\\
 &  & p_{0}+p_{s}+\sum_{k=1}^{K}p_{k}\leq P,
\end{eqnarray*}
where $R(p_{0},p_{s},\{p_{k}\})$, $\widehat{\mathrm{CRB}}(\theta_{s},p_{0},p_{s},\{p_{k}\})$,
and $\widehat{\mathrm{CRB}}(r_{s},p_{0},p_{s},\{p_{k}\})$ are the
corresponding formulas after variable change. Problem (P4) can be
optimally solved via a similar approach as for (P1), for which the
details are omitted for brevity.

\section{Numerical Results}

In this section, we provide numerical results to validate the effectiveness
of our proposed near-field joint secure beamforming designs for the
ISCAP system. We assume that the BS is equipped with $N=64$ antennas
and the carrier frequency is set as $3\textrm{ GHz}$ such that $\lambda=0.1\textrm{ m}$.
Consider half-wavelength spacing, we have $d=0.05\textrm{ m}$ and
the Rayleigh distance is around $198\textrm{ m}$. To better illustrate
the beamforming performance in the angle and distance domain, we adopt
the near-field LoS channel model. Furthermore, we set the angles of
CU and target to be identical, i.e., $\theta_{s}=\theta_{0}=60{^\circ}$,
the distance of CU is set as $r_{0}=5\textrm{ m}$, and $K=2$ ERs
and randomly located in the angle region $[90{^\circ},120{^\circ}]$
and range region $[5\textrm{ m},8\textrm{ m}]$, The total transmit
power is set as $P=43\textrm{ dBm}$. Furthermore, the CRB thresholds
for angle and distance are set as $\Gamma_{\theta}=\Gamma_{r}=0.1$
and the harvested power threshold is set as $Q=0.025\textrm{ mW}$.
The noise power is set as $\sigma_{0}^{2}=\sigma_{k}^{2}=-40\textrm{ dBm}$.
For comparison, we consider a benchmark design based on separate beamforming,
in which the sensing/energy covariance $\boldsymbol{S}$ is first
designed with a minimum power to satisfy the CRB and energy harvesting
requirements. Then, the information transmit beamforming $\boldsymbol{w}_{0}$
is designed to achieve the maximum secrecy rate. 
\begin{figure}

\vspace{-0.25cm} \includegraphics[scale=0.45]{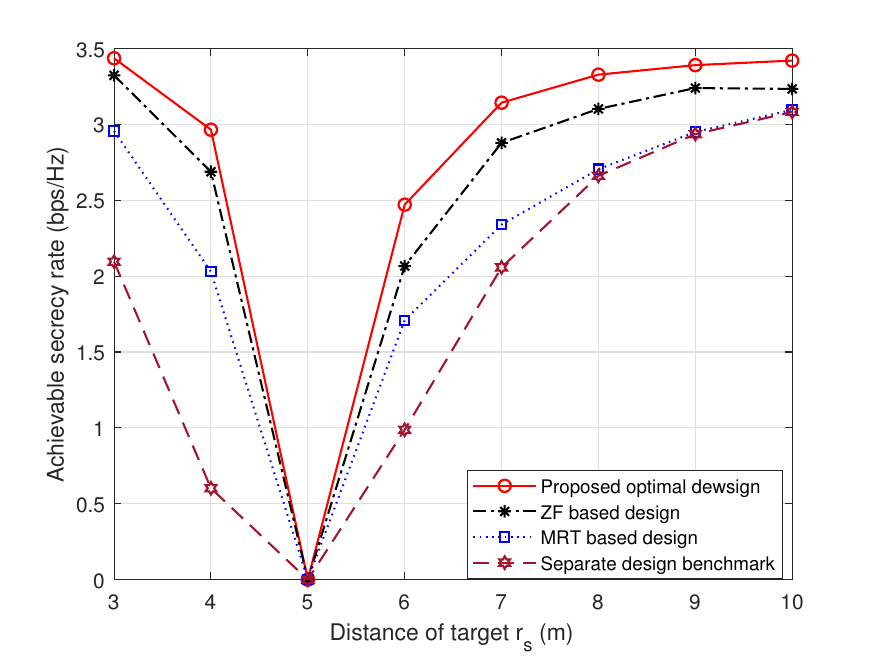}\vspace{-0.2cm}\centering\caption{Achievable secrecy rate versus the target distance $r_{s}.$}

\end{figure}

Fig. 2 shows the achievable secrecy rate versus the target distance
$r_{s}$. It is observed that the achievable secrecy rate first decreases
to zero at distance $r_{s}=5\textrm{ m}$, then increases with the
distance $r_{s}$. It is shown that although the CU and the target
are located in the same direction with respective to the BS, non-zero
secrecy rate is still achievable via exploiting the difference in
the distance domain in near-field scenarios. This is in sharp contrast
to the far-field beam steering. It is also observed that these two
alternative designs achieve satisfactory secrecy rates comparable
to the optimal design and outperforms the separate design benchmark.
This shows the effectiveness of these designs.

\begin{figure}
\vspace{-0.28cm} \includegraphics[scale=0.45]{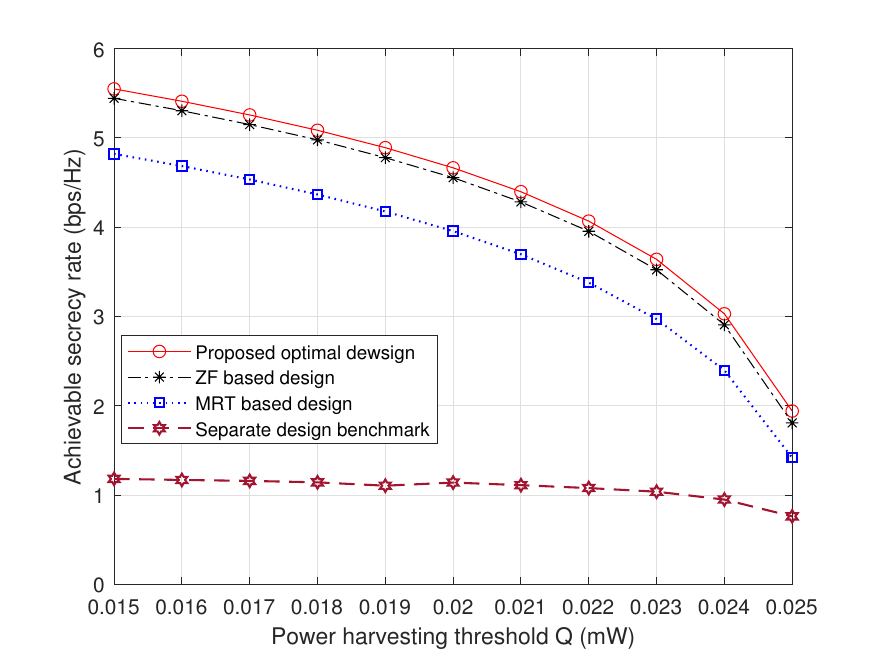}\vspace{-0.2cm}\centering\caption{Achievable secrecy rate versus the power harvesting threshold $Q$.}
\vspace{-0.4cm} 
\end{figure}

Fig. 3 shows the achievable secrecy rate versus the power harvesting
threshold $Q$, with the sensing target located at a distance of $r_{s}=6\textrm{ m}$.
It is observed that the optimal design achieves the best performance
among all schemes, and ZF-based design demonstrates superior secrecy
performance compared to the MRT-based design. This is due to the fact
that with $N=64$ in this case, there are sufficient design degrees
of freedom for implementing ZF beamforming to achieve satisfactory
secrecy rate performance.

\section{Conclusion}

This paper investigated a secure ISCAP system with one ELAA-BS serving
one single CU, one single sensing target, and multiple ERs, where
both the target and ERs are potential eavesdroppers. We proposed a
novel joint information and sensing/powering/AN beamforming design
to maximize the secrecy rate while ensuring the perfromance requirements
on WPT and target sensing. We proposed the optimal solution based
on the SDR and fractional programming techniques together with 1D
search. Numerical results were provided to demonstrate the effectiveness
of our proposed methods. It is shown that our proposed approaches
utilized near-field ELAA's distance-domain resolution and joint beamforming
to enhance secure communication in ISCAP.

{\footnotesize{}{} \bibliographystyle{IEEEtran}
\bibliography{IEEEabrv,IEEEexample,myref}
}\vspace{-0.1cm}{\footnotesize{} }{\footnotesize\par}

\end{document}